\documentclass[amsmath,amssymb,aps,twocolumn,prb,freqn,floatfix]{revtex4-2}
\usepackage{graphicx}
\usepackage{dcolumn}
\usepackage{bm}
\usepackage{hyperref}
\usepackage{color}
\usepackage{subfigure}
\usepackage{flexisym}

\begin{document}

\title{Realising Einstein's mirror: Optomechanical damping with a thermal photon gas}

\author{A T M Anishur Rahman}
 \altaffiliation{Department of Physics and Astronomy,
 University College London, WC1E 6BT, London, UK}
 \email{a.rahman@ucl.ac.uk}
\author{P. F. Barker}
\altaffiliation{Department of Physics and Astronomy, University College London, WC1E 6BT, London, UK}
\email{p.barker@ucl.ac.uk}
%

\date{\today}

\begin{abstract}
In 1909 Einstein described the thermalization of a mirror within a blackbody cavity by collisions with thermal photons. While the time to thermalize the motion of even a microscale or nanoscale object is so long that it is not feasible, we show that it is using the high intensity light from an amplified thermal light source with a well-defined chemical potential. We predict damping of the center-of mass motion due to this effect on times scales of seconds for small optomechanical systems, such as levitated nanoparticles, allowing experimental observation. 
\end{abstract}

\maketitle
In 1909 Einstein described how an object's motion would be damped by light scattering processes when placed inside a blackbody (BB) cavity \cite{Einstein1909}. Here, in analogy with Brownian motion, a dynamic equilibrium between the momentum fluctuations of the BB light and the object would bring the motional temperature to that of the blackbody. Importantly, both the wave-like nature of the radiation via interference processes and the particle-like nature of the photons contribute to this process and was used to understand the Planck description of the properties of blackbody sources. This process was further explored as a potential mechanism for damping on astronomical scales \cite{HeerKohlPR1968,HenryPetersonPR1968} by thermal radiation pressure from the cosmic microwave background at 3 K. However, it was shown that the damping time for any object was significantly longer than the lifetime of the universe and therefore would not be of significance in astrophysical processes \cite{HeerKohlPR1968}. This damping process is weak, because as the temperature of a BB decreases, the number of photons per unit volume decreases. This occurs as a blackbody has a chemical potential of zero. In addition, the large spectral range of a blackbody means that it would be difficult to focus all the light on to a small object such as a mirror.

Thermal sources of light with a well defined chemical potential have only recently been realised. These sources allow control over the chemical potential and therefore the number of photons per unit volume for a fixed temperature \cite{KlaersNatPhys2010,KlaersNat2010,WeillRafiNatComm2019,MarelicPRA205}. These thermal sources are in a dynamic equilibrium where the photons come into thermal equilibrium with an active medium via absorption and emission. Since they can be pumped optically, and the photons can be confined within a cavity, the chemical potential can be varied such that even Bose-Einstein condensation has been realized \cite{KlaersNat2010,WeillRafiNatComm2019}. In addition, the limited spectral range of the resonant transitions of these sources means that the radiation is relatively narrowband ($<$ 100 nm) when compared to a BB source (1000s nm at 300 K) and therefore, unlike a blackbody, practically all of the light can be focused and used to interact optomechanically with an object, such as a mirror or a dielectric nanoparticle.

Here, we show that such light sources could be used to cool and damp the motion of nanoscale levitated particles held in a trap, which in the absence of additional feedback or cavity cooling, would heat via recoil of laser photons to motional temperatures in excess of 1000s of Kelvin. \cite{NovotnyPRA2017}. 

We first outline the thermalisation of the motion of a mirror to a blackbody photon gas. While Einstein and others considered the object placed inside the blackbody \cite{Einstein1909,Mansuripur2017}, we consider the mirror outside the cavity illuminated by light emanating from the wall of a $3D$ blackbody cavity. The mirror is perfectly reflecting at all wavelengths and is a disk of area $A$ and mass $M$. The spectral distribution of the photons is described using the Bose-Einstein (BE) distribution. We consider that the disk is in motion with a velocity $v_z$ along the $z~$ axis. The photons make an angle $\theta$ with the surface normal of the disk in the $z$ direction. The number density and the variance of the blackbody photons per unit angular frequency, per solid angle and per volume are given by $\rho_n=\frac{\omega^2}{4\pi^3c^3}~\frac{1}{\exp{(\hbar\omega/k_BT)}-1}$ and $\Delta N^2=\frac{\omega^2}{4\pi^3c^3}~\frac{\exp{(\hbar\omega/k_BT)}}{\bigl[\exp{(\hbar\omega/k_BT)}-1\bigr]^2}$, respectively \cite{Mansuripur2017}. In the reference frame of the disk, the frequency of the incident radiation appears shifted due to the Doppler effect. To the disk in the moving frame, the BB radiation has an effective temperature \cite{HeerKohlPR1968,PeeblesPR1968} $T (1+\beta_z\cos\theta)^{-1}$ and the number density of photons per unit angular frequency and solid angle in this frame is  $\rho_n(v_z)=\frac{\omega^2}{4\pi^3c^3}~\frac{1}{\exp{(\hbar\omega(1+\beta_z\cos\theta)/k_BT)}-1}$, where $\beta_z=v_z/c$ and that $v_z\ll c$. The total momentum that is delivered to a disk of area $A$ that is illuminated by the light from a BB (with solid angle $\Omega=2 \pi$) leads to a total force \cite{HeerKohlPR1968,PeeblesPR1968}
\small
\begin{eqnarray}
\nonumber
F_z(v_z)&=&\int_0^{\pi/2}\int_0^{2\pi}\int_0^{\infty} c A \cos\theta \rho_n(v_z) 2\hbar k\cos\theta~d\Omega d\omega\\
&\approx&\frac{A\pi^2k_B^4T^4}{45c^3\hbar^3}-\frac{A\pi^2k_B^4T^4}{15c^4\hbar^3}v_z,
\label{eqn1}
\end{eqnarray}
\normalsize
where $d\Omega=\sin\theta d\theta d\phi$. The first term in equation (\ref{eqn1}) is the usual radiation pressure force while the second term is radiation damping with a rate given by $\Gamma_z=\frac{A\pi^2k_B^4T^4}{15Mc^4\hbar^3}$. The amount of energy that the disk loses per second is $(M\Gamma_zv_z)~v_z=M\Gamma_zv_z^2$, where from the kinetic theory $v_z^2=k_BT_{cm}/M$ \cite{Mansuripur2017} and $T_{cm}$ is the centre-of-mass (CM) temperature of the disk. In addition, due to the fluctuation in the photon number, and the associated momentum kicks from the impinging photons \cite{Einstein1909,Mansuripur2017}, the energy that the disk gains in unit time is 
\small
\begin{eqnarray}
\nonumber
\Delta \dot E&=&\frac{1}{2M}\int_0^{\pi/2}\int_0^{2\pi}\int_{0}^{\infty}{cA\cos\theta\Delta N^2(2\hbar k\cos\theta)^2 d\Omega d\omega}\\
&=&\frac{A\pi^2k_B^5T^5}{15M\hbar^3c^4}.
\label{eqn2}
\end{eqnarray}
\normalsize

At equilibrium, the rate in loss and the rate in gain in energy are equal so that $M\Gamma_zv_z^2=\Delta \dot E$, and  the centre-of-mass temperature is found to be equal to the blackbody temperature.

\begin{eqnarray}
T_{cm}^z&=&T.
\label{eqn3}
\end{eqnarray}
 This was discussed by Einstein \cite{Einstein1909} and later calculated for inside the 3K cosmic blackbody radiation of the universe \cite{HeerKohlPR1968,HenryPetersonPR1968}.
 
\begin{figure}
    \centering
    \includegraphics[width=8.5cm]{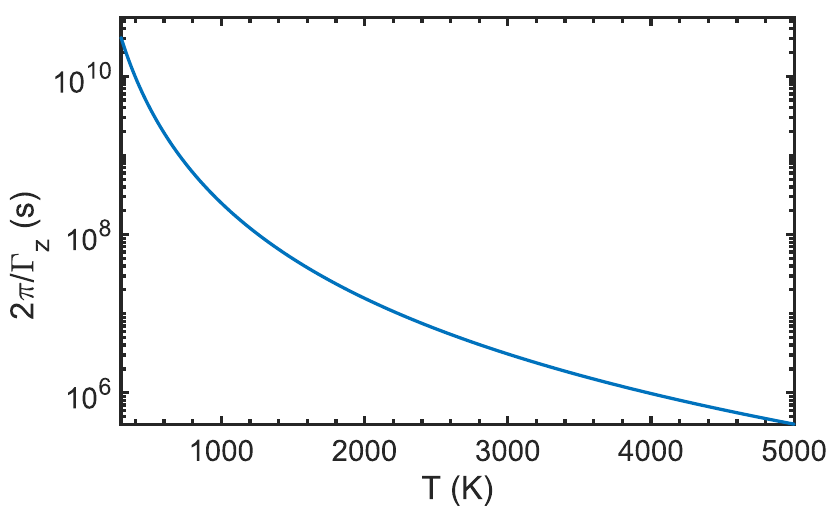}
    \caption{The radiation damping time, $2\pi/\Gamma_z$, as a function of the blackbody temperature calculated for a circular silica disk of radius $r=5~\mu$ m, thickness $50~$nm and density 2000~kg~m$^{-3}$. This is equivalent to a mass $M\approx 7.85\times 10^{-15}$~kg.}
    \label{fig2}
\end{figure}

To demonstrate how weak this damping process is for even a hot BB source we consider the damping of a silica disk of radius $r=5~\mu$m, thickness $50~$nm, density 2000~kg~m$^{-3}$ and mass of $M\approx 7.85\times 10^{-15}~$kg. The best case damping time, where unrealistically the maximum $2 \pi$ solid angle of the BB source could be captured and focused to the size of the disk, is given by $\tau_z=2\pi/\Gamma_z$. A plot of the damping time is shown as a function of the BB source temperature in Fig. \ref{fig2}. The damping time decreases rapidly as the temperature decreases. This is due to the decrease in photon flux with blackbody temperature($\approx T^4$). This occurs as a blackbody has no chemical potential and the photon number cannot be conserved as the temperature changes \cite{Wurfel_1982}. At a BB temperature of $300~$K, the damping time for the silica disk is approximately $1000$ years. At a temperature of $5000~$K the damping time reduces to $\approx 4.63~$days. While this time is significantly less than that at room temperature, it is still at the limit of experimental verification \cite{PontinPRR2020}.  Lastly, producing such a high temperature blackbody source would be challenging within a laboratory environment. 

\textit{ Damping of a reflecting disk by a 2D thermal photon gas:} Over the last decade new thermal light sources with non-zero chemical potentials have been realised. Bose-Einstein condensates of photons have been achieved by increasing the chemical potential by strongly pumping these systems  \cite{KlaersNatPhys2010,KlaersNat2010,WeillRafiNatComm2019}. These optical sources have been produced by optical pumping of cavities containing dyes in liquids or rare earth ions within fibres. They are operated below the lasing threshold and the photons come into thermal equilibrium with the matter. An important property of these sources is that unlike a blackbody source, the chemical potential and therefore the photon flux, can be controlled or maintained when the temperature is changed. This opens up the possibility of producing more intense sources per unit frequency when compared to thermal radiation produced by blackbody sources. Coupled with the ability to isolate microscopic particles from environmental heating sources using optomechanical methods, we show that these sources should allow the experimental realisation of optical damping due to the thermal nature of light as envisioned by Einstein.

 \begin{figure*}[t]
 	\centering
 	\includegraphics[width=17.50cm]{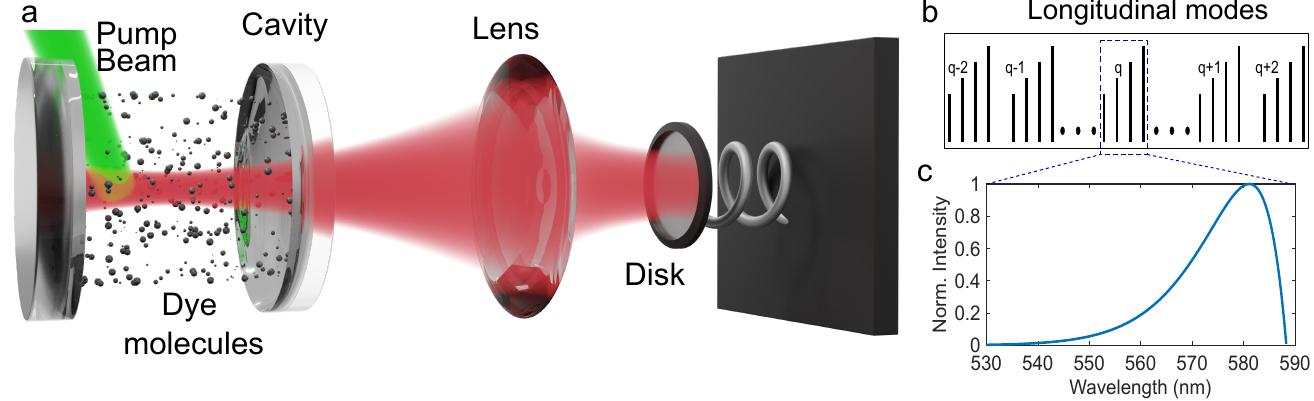}
 	\caption{\textbf{A schematic diagram of damping by thermal radiation~} a) A micro-cavity containing dye molecules forms a thermalized photon gas. Multiple absorption-emission cycles in the dye molecules provide the necessary thermalization of photons while the cavity traps photon facilitating thermalization providing a well defined set of transverse modes. For a 2-D cavity, only a single longitudinal cavity mode is occupied, while thermalisation occurs in the transverse modes. The number of dye molecules inside the cavity determines the chemical potential. Thermalized photons emitted from the cavity mirrors are collected by a lens and focused onto an ideal perfectly reflecting disk. On reflection, a thermalized photon delivers a momentum kick proportional to its wavenumber and the translational velocity of the disk. The light is emitted along the $z-$axis while the disk is in the $x-y~$plane. b) Thermalisation in transverse modes for a single transverse mode. The emission spectrum of the dye molecules is determined by the longitudinal and transverse cavity mode occupied and the spectral profile of the dye. c) The spectral profile of the 2-D cavity. The Bose-Einstein spectral density of a single longitudinal mode $q$, where we have assumed $2\pi c/\omega_c=588.24~$nm and $\mu_c=1.93~$eV \cite{KlaersNatPhys2010}.}
 	\label{fig0}
 \end{figure*}
As a concrete example, we consider a 2D microcavity consisting of two cavity mirrors filled with dye molecules (see Fig. \ref{fig0}) \cite{KlaersNatPhys2010}. Here, the cavity works as a trap for the photons emitted by the dye molecules when optically pumped. In addition, the dye molecules act as a thermal bath for the photons and provide the necessary chemical potential required for the conservation of the number of photons when the ambient temperature is varied \cite{KlaersNatPhys2010}. The photon statistics of these sources are still given by the usual Bose-Einstein distribution with the inclusion of a chemical potential $\mu_c$ \cite{KlaersNatPhys2010,Sobyanin2013}. The energy density inside the cavity is given by 
\small
\begin{eqnarray}
\nonumber
\bar{u}&=&\frac{1}{V_R}\sum_{n_x=0}^{\infty}\sum_{n_y=0}^{\infty}2(n_x+n_y+1)\\
&\times&~\frac{\hbar(\omega_c+(n_x+n_y+1)\Omega) }{\exp{[(\hbar(\omega_c+(n_x+n_y+1)\Omega)-\mu_c) /k_BT]}-1},
\end{eqnarray}
\normalsize
where $n_x$ and $n_y$ are the transverse mode number, $\Omega=2\pi c/\sqrt{D_0R/2}$ is the difference in frequency between two consecutive transverse modes, $\omega_c=q \pi c/D_0$ is the angular frequency of the longitudinal mode number $q$, $D_0$ is the spatial separation between the two cavity mirrors, $R$ is the radius of curvature of the cavity mirrors, $n$ is the refractive index of the dye medium, and $V_R$ is the volume of the cavity. In the continuum limit ($\Omega\rightarrow0$) \cite{MullerEPRA2019}, the average number of photons that is transmitted through one of the mirrors of the high finesse cavity (see supplementary information for details), per angular frequency and solid angle, is now $\dot{\bar{N}}=\frac{V_RT_r}{n q D_0}\frac{\omega_c\omega}{4\pi^3c^2}\frac{1}{\exp{[\hbar(\omega_c+\omega)-\mu_c)/k_BT]}-1}$, where $T_r$ is the transmission co-efficient of the cavity mirror. Given that $\exp{[\hbar(\omega_c+\omega)-\mu_c)/k_BT]}\gg 1$ \cite{KlaersNatPhys2010}, $\dot{\bar{N}}$ can be approximated as $\frac{V_RT_r}{qnD_0}\frac{\omega_c\omega}{4\pi^3c^2}\frac{1}{\exp{[\hbar(\omega_c+\omega)-\mu_c)/k_BT]}}$. Output powers in the range of $10$s of nano-watts have been demonstrated where the power and chemical potential can be varied via by the optical pumping power and by the number density of the dye molecules \cite{KlaersBECTutorial2014}. Importantly, due to the relatively narrow bandwidth ($\approx 60~$nm \cite{KlaersNatPhys2010}) of the light compared to a blackbody source, this light can be amplified further using optical amplifiers with gain, $G$, to increase the power while maintaining the photon statistics. As the beam can be tightly focused using a microscope objective, the intensity can be orders of magnitude higher than that for even a very hot blackbody source. When illuminated with such a light, the force that a perfectly reflecting disk of area $A$ encounters in the moving frame is
\small
\begin{eqnarray}
\nonumber
F_z&=& \frac{G V_RT_r \exp{[\mu_c/k_BT]}}{qnD_0}\\
\nonumber
&\times& \int_0^{\frac{\pi}{2}}\int_0^{2\pi} \int_0^{\infty}{ \frac{\omega_c \omega }{4\pi^3 c^3} \frac{2\hbar(\omega_c+\omega)\cos^2\theta d\Omega d\omega }{\exp{[(\hbar(\omega_c+\omega))(1+\beta \cos{\theta}) /k_BT]}}}\\
\nonumber
&\approx &\frac{G V_RT_r }{qnD_0}  \frac{\exp{[(\mu_c-\hbar\omega_c)/k_BT]}\hbar^2\omega_c^2k_B^2T^2}{3\pi^2 \hbar^3 c^3}\\
&&-\frac{G V_RT_r }{qnD_0}  \frac{\exp{[(\mu_c-\hbar\omega_c)/k_BT]}\hbar^3\omega_c^3k_BT }{4\pi^2 \hbar^3c^4}~v_z,
\label{eqn4}
\end{eqnarray}
\normalsize
where we have assumed that the diameter of the incident light beam (see Fig. \ref{fig0}) is equal to or smaller than that of the disk. The corresponding damping rate is 
\begin{eqnarray}
\Gamma_z&=&\frac{G V_RT_r }{4q D_0M}  \frac{\exp{[(\mu_c-\hbar\omega_c)/k_BT]}k_BT \omega_c^3}{\pi^2c^4}.
\label{eqn41}
\end{eqnarray}
The rate of energy gain due to the fluctuating photon momentum is now 
\begin{equation}
  \Delta \dot E_z\approx\frac{G V_RT_r }{4q D_0M}  \frac{\exp{[(\mu_c-\hbar\omega_c)/k_BT]}k_B^2T^2 \omega_c^3}{\pi^2 c^4}. 
  \label{eqn5}
\end{equation} 

\begin{figure}
    \centering
    \subfigure{\includegraphics[width=8.25cm]{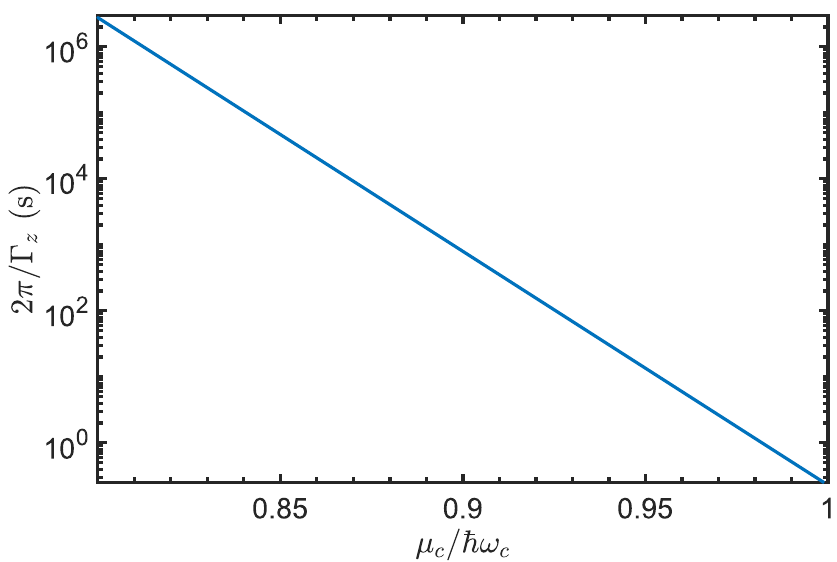}}
    \caption{The relaxation time, $2\pi/\Gamma_z$, as a function of the chemical potential normalized by the cavity cut-off frequency. The cavity has a mirror transmission $T_r=1.5\times10^{-5}$, $\lambda_c=2\pi c/\omega_c=588~$nm, longitudinal mode number $q=7$,  mirror separation of $D_0=1.45~\mu$ m and radius of curvature $R=1~$m. The disk has radius $5~\mu$ m, thickness $50~$nm. The thermal source is at $T=300~$K and an optical gain of $G=80~$dB.}
    \label{fig3}
\end{figure}
The equilibrium centre-of-mass temperature of the mirror is $T_{cm}=\Delta\dot{ E_z}/k_B\Gamma_z=T$. This is the same as that obtained from a blackbody source. However, importantly now, both $\Gamma_z$ and $\Delta \dot E_z$ are adjustable through the chemical potential $\mu_c$ and the optical gain $G$. Figure \ref{fig3} shows $2\pi/\Gamma_z$ as a function of normalized chemical potential $\mu_c/\hbar\omega_c$ at $T=300~$K. The parameters used in the calculation are typically used in 2D experimental microcavities \cite{KlaersNatPhys2010}. For a chemical potential $\mu_c/\hbar\omega_c=0.92$, and an optical amplifier gain of $80~$dB giving an output power of $200~$mW, we calculate a damping time of $2\pi/\Gamma_z\approx 150~$ seconds. This is six orders of magnitude larger than from a blackbody source at 4000 K. 

For comparison, we consider the same reflective disk considered above, but now illuminated by a laser beam with the same intensity and whose frequency is equal to the cutoff frequency of the 2-D cavity. If this system can come into equilibrium via radiation, then the damping rate is $\Gamma_z=2\dot{\bar{N}}\hbar\omega_c/(Mc^2)$, and the corresponding rate of gain in energy is $\Delta \dot{E_z}=2\bar{N}\hbar^2\omega_c^2/Mc^2$, where $\omega_c$ is the angular frequency of the laser and $\dot{\bar{N}}$ is the average photon number per unit time in the laser beam. The variance of a Poisson distributed laser beam is also $\dot{\bar{N}}$. Equating these two quantities gives a centre-of-mass temperature of the disk of $\hbar\omega_c/k_B$, which is $24461~$K for a $\lambda_c = 588~$nm laser. This is significantly higher than the thermal source at 300 K.

\begin{figure*}[t]
    \centering
    \includegraphics[width=17.50cm]{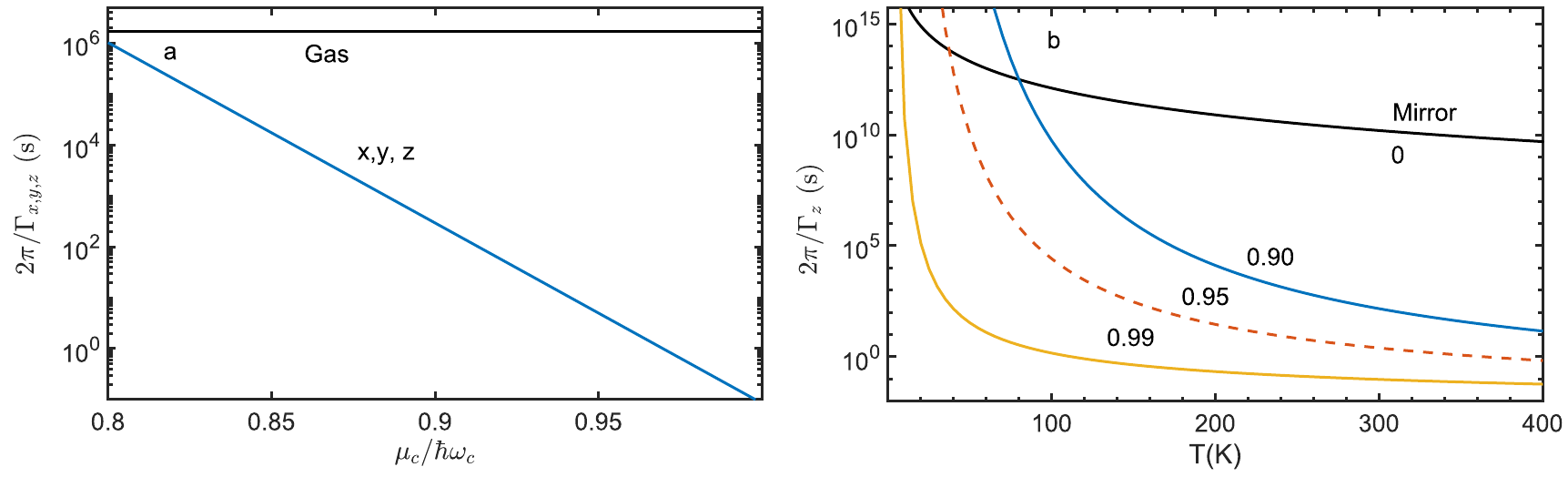}
    \caption{\textbf{Radiation damping times for a levitated silica sphere} - \textbf{a)} The damping time ($2 \pi/\Gamma_{x,y,z}$) of a $r=100~$ nm silica sphere along the x, y and z axes when illuminated with a thermal photon gas propagating along the z axis and focused to a spot size of $w=1~\mu$m. The relaxation time is calculated for light produced by a 2-D cavity and amplified by $G=70~$dB. The separation between the cavity mirrors is $D_0=1.5~\mu$ m with a transmission co-efficient of $T_r=1.5\times10^{-5}$, and $2q\pi c/\omega_c=588~$nm \cite{KlaersNatPhys2010}. For $\mu_c/\hbar\omega_c=0.92$, the equivalent optical power after amplification is $200~$mW or $6.4\times 10^{10}~$W~m$^{-2}$ at the location of the particle. For comparison, we have also included the damping time to reach the equilibrium  associated with the collisions with the background gas molecules. In this case, the background gas pressure is $1\times 10^{-9}~$mBar. \textbf{b)} The relaxation time of the sphere of part a) along the $z-$axis as a function of temperature for three different chemical potentials- $0.99, ~0.95,~ \&~ 0.90$. For comparison, we have also shown the relaxation time of a $r=100~$nm disk when illuminated with a thermal blackbody source with $\mu_c=0$.}
    \label{fig4}
\end{figure*}

\textit{A levitated dielectric sphere damped by a thermal photon gas:} Finally, we consider a levitated dielectric sphere of radius $r\ll 2\pi c/\omega_c$ and a scattering cross section of $\sigma_s=\frac{\alpha^2\omega_i^4}{6\pi\epsilon_0^2c^4}$ is illuminated by an amplified 2D thermal light from a microcavity, where the frequency of the incident light field is $\omega_i=\omega_c+\omega$ and $\alpha$ is the polarizability of the particle\cite{BohrenHuffman2007}. Furthermore, we assume that the amplified light is tightly focused using a lens to a spot size of area $A_w$. The particle could be levitated in a Paul trap, or by the thermal light itself with sufficient intensity. For simplicity we assume that the scattering cross section of the particle is independent of velocity. In the laboratory frame, the wavevector of an incident photon is given by $\mathbf{k}_i=[k_x~k_y~k_z]$, where $k_i=\omega_i/c$. In contrast, in the reference frame of the particle the frequency of an incident and a scattered photon appear as $\omega_i(1+\beta)$, where $\beta=\frac{\mathbf{v}\cdot\mathbf{k}_i}{ck_i}$ and $\mathbf{v}=[v_x ~v_y~v_z]$ is the velocity of the particle along the three translational axes. Owing to the unpolarized nature of the thermal incident photons, the scattered photons are isotropically distributed over $4\pi$ steradians. After a scattering event each photon delivers a momentum to the particle equivalent to $\mathbf{p}=\hbar k_i(1+\beta)(\mathbf{\Theta}_i-\mathbf{\Theta}_s)$, where $\mathbf{\Theta}_i=[\sin\theta_i\cos\phi_i~\sin\theta_i\sin\phi_i~\cos\theta_i]$ and $\mathbf{\Theta}_s=[\sin\theta_s\cos\phi_s~\sin\theta_s\sin\phi_s~\cos\theta_s]$. Here, $\theta_i$ and $\phi_i$ represent the polar and the azimuthal angles that the wavevector of an incoming photon makes with the $-z~$axis and the $+x~$axis, respectively. $\theta_s$ and $\phi_s$ are the corresponding angles that the wavevector of a scattered photon makes with the same reference axes. With the appropriate Lorentz transformation, the total force exerted by all photons that interact with the particle is
\begin{eqnarray}
\nonumber
\mathbf{F}&=&\frac{ GV_RT_r}{q nD_0}~\frac{1}{4\pi}\int_0^\pi\int_0^{2\pi}\int_0^{\frac{\pi}{2}}\int_0^{2\pi}\int_{0}^{\infty}\Bigg[\frac{\sigma_{s}}{A_w}\frac{\omega_c\omega }{4\pi^3c^2}\\
\nonumber
&&\times \frac{\exp{(\mu_c/k_BT)}}{\exp{\Bigl[\hbar\omega_i(1+\beta)/k_BT\Bigr]}}~ \hbar k_i\Bigl(\mathbf{\Theta}_i -\mathbf{\Theta}_s\Bigr) \Bigg]d\omega d\Omega_id\Omega_s \\
&\approx&- \frac{GV_RT_r}{qnA_wD_0}~\frac{\exp{[(\mu_c-\hbar\omega_c)/k_BT]}\alpha^2k_BT\omega_c^7}{36\pi^3 \epsilon_0^2c^8}~\mathbf{v}
\end{eqnarray}
\normalsize

where $d\Omega_i=\sin\theta_i  d\phi_i d\theta_i$ with $\theta_i^{max}=\pi/2$ and $\phi_i^{max}=2\pi$, and $d\Omega_s=\sin\theta_s  d\phi_s d\theta_s$ with $\theta_s^{max}=\pi$ and $\phi_s^{max}=2\pi$. In the final result, we have only shown the velocity dependent term. The detail derivation is available in the supplementary information. The damping rate along the three axes is given by $\Gamma_{x,y,z}=\frac{GV_RT_r}{qnA_wD_0M}~\frac{\exp{[(\mu_c-\hbar\omega_c)/k_BT]}\alpha^2k_BT\omega_c^7}{36\pi^3 \epsilon_0^2c^8}$. Figure \ref{fig4}a shows the time required by a $r=100~$nm silica sphere to reach equilibrium from an arbitrary state. Due to the isotropic nature of scattering, the particle requires the same time to reach the equilibrium along all three axes. At $\mu_c/\omega_c=0.92$, the required time for the particle to reach the equilibrium is $2\pi/\Gamma_{x}\approx 60~$seconds. This time can additionally be controlled by using a larger or a smaller sphere. In ultra high vacuum ($10^{-9}~$mBar), the thermalisation time due to collisions with the residual air molecules, is $1.72\times 10^6~$sec. This is approximately five orders of magnitude larger than that from the thermal photons. Figure \ref{fig4}b shows the effect of the change in cavity bulk temperature on the thermalization time. For a fixed chemical potential $2\pi/\Gamma_z$ increases rapidly as the temperature goes down. However, this can be counteracted by increasing the pump power for a fixed dye molecule number density or by increasing the dye molecule density for a fixed pump power \cite{KlaersNat2010,KlaersNatPhys2010}. 

The gain in kinetic energy by the particle due to the fluctuation in momentum \cite{Einstein1909,SebersonPRA2020,ItanoWinelandPRA82} is

\small
\begin{eqnarray}
\nonumber
\Delta \dot E&=&\frac{1}{2M}\frac{ GV_RT_r}{q nD_0}~\frac{1}{4\pi}\int_0^\pi\int_0^{2\pi}\int_0^{\frac{\pi}{2}}\int_0^{2\pi}\int_{0}^{\infty}\Bigg[\frac{\sigma_{s}}{A_w}\frac{\omega_c\omega }{4\pi^3c^2}\\
\nonumber
&&\times \frac{\exp{(\mu_c/k_BT)}}{\exp{\Bigl(\hbar\omega_i/k_BT\Bigr)}} ~\hbar^2 k_i^2\Bigl(\mathbf{\Theta}_i -\mathbf{\Theta}_s\Bigr)^2 \Bigg]d\omega d\Omega_i d\Omega_s \\
&\approx& \mathbf{\Lambda} \frac{GV_RT_r}{qnA_wD_0M}~\frac{\exp{[(\mu_c-\hbar\omega_c)/k_BT]}\alpha^2k_B^2T^2\omega_c^7}{36\pi^3 \epsilon_0^2c^8},
\end{eqnarray}
\normalsize
where $\mathbf{\Lambda}=[1~1~1]$. The centre-of-mass temperature of the particle along all three axes is equal to the bulk temperature $T_{cm}=T$. This is strikingly different than that is found \cite{NovotnyPRA2017} for a laser illuminated spherical particle  $T_{cm}=\hbar\omega_c/4k_B$ or $6115~$K for a $2\pi c/\omega_l=588~$nm laser and can be used to reduce the effects of recoil heating in levitated experiments such that no feedback cooling is required to keep a particle in an optical trap.

For a $r=100~$nm sphere in UHV ($10^{-9}~$mBar), the damping rate due to the gas molecules is $\approx 5.81\times 10^{-7}~$Hz - about five orders of magnitude less than that is exerted by the thermalized photon gas. This means that the radiation damping should be easily detectable. For an actual measurement of the radiation damping encountered by a levitated object, the levitated object can be excited (cooled) to a higher (lower) energy state, for example by manipulating the trapping potential \cite{GieselerNatNano2014,RahmanOptica2020} or by an electric field if it is charged \cite{TebbenjohannsPRL2019}, followed by a ring-down (reheating) style measurement for determining $\Gamma$.

We have shown that small, well isolated optomechanical systems, such as those produced by levitation, could be damped and thermalized with a thermal photon gas as originally envisioned by Einstein in 1909. Such an experiment is feasible using thermalized light sources, \cite{KlaersNatPhys2010,WeillRafiNatComm2019}, and recent advances in optomechanics. Here, for an experimental demonstration, a levitated optomechanical object such as a charged nanoparticle in a Paul trap \cite{PontinPRR2020} or a neutral nanoparticle in an optical trap \cite{RahmanSciRep2016,RahmanNatPhot2017,RahmanRSI2018, Gieseler2012,Vovrosh2017,GambhirPRA2016} in ultra high vacuum seems ideal. Although we have calculated the damping from a 2-D source, a 1-D source which has only longitudinal modes could be focused more tightly which would lead to higher damping than shown here. Light sources such as amplified LEDs and superluminescent diodes which produce thermal light \cite{Wurfel_1982,Hartmann2017}, and have been used to trap dielectric spheres \cite{RahmanOptica2020}, could also be considered for investigating thermal radiation damping. However, the spectral profile of these devices is not as well defined as in the cavity sources studied here and would require further modelling. Finally, our results raise the possibility that by increasing/decreasing the bulk temperature of a thermalized light source one can heat/cool the centre-of-mass temperature of an optomechanical system which is currently controlled by using either parametric feedback cooling \cite{RahmanOptica2020,Vovrosh2017,Gieseler2012} or velocity damping \cite{TebbenjohannsPRL2019,GambhirPRA2016}. 


%

\clearpage
\LARGE


\clearpage
\widetext
\appendix

\LARGE
\textbf{Supplementary information}
\normalsize

\section{Blackbody temperature in a moving frame.}
\label{supA0}
Consider a stationary frame $s$ and an object moving in frame $s'$ with a velocity $v$. For simplicity, we assume that all axes in these two frames are aligned. In this case, quantities in the $s'$ frame are related with those in the $s$ frame through the following identities of the Lorentz transformation between the two frames \cite{PeeblesPR1968}
\begin{eqnarray}
\nonumber
dt'&=&dt/\gamma  \\
\nonumber
\omega'&=&\gamma  (1+\beta \cos{\theta})\omega\\
\frac{d\Omega'}{d\Omega}&=&[\gamma (1+\beta\cos\theta)]^{-2}\\
\nonumber
 \cos\theta'&=&\frac{\cos\theta+\beta}{1+\beta\cos\theta}\\
\nonumber
\label{eqnA0}
\end{eqnarray}
where $\omega$ and $\omega'$ are the angular frequencies of the light in each frame, $\gamma=1/\sqrt{1-v^2/c^2}$, $\beta=v/c$, and $\Omega$ and $\Omega'$ are the solid angles in each frame. The angle $\theta$ and $\theta'$ are the angles between the surface normal of the photon detector and  the  wavevector  of an  incident  photon in each frame. 

 The number of photons that a surface of area $A$ in the frame $s'$ receives in time $dt'$ from a blackbody source that is stationary in frame $s$ is
\begin{eqnarray}
dN'=\rho_n' d\omega' d\Omega' A c \cos \theta'  dt',
\end{eqnarray}
where $\rho_n'$ is the number of photons per unit volume per unit angular frequency per unit solid angle given by
\begin{eqnarray}
\rho_n'(\omega',\theta')=\frac{\omega^{'2}}{4\pi^3 c^3}\frac{1}{\exp{[\hbar\omega'/k_B  T']}-1}.
\end{eqnarray}
From the point of view of a stationary observer \cite{PeeblesPR1968}, the same photons are represented in the moving frame as
\begin{eqnarray}
dN=\rho_n d\omega d\Omega A c (\beta+ \cos\theta)dt.
\end{eqnarray}
As the number of photons is equal in both frames, $dN'=dN$, then
\begin{eqnarray}
\rho_n' d\omega' d\Omega' A c \cos \theta' dt' &=& \rho_n d\omega d\Omega A c (\beta+ \cos\theta)dt\\
\rho_n'&=&\gamma^2(1+\beta\cos\theta)^2~\rho_n
\label{eqnA1}
\end{eqnarray}
Now substituting the photon number density \cite{Mansuripur2017} in two different frames we have,
\begin{eqnarray}
\nonumber
\frac{\omega^{'2}}{4\pi^3c^3}\frac{1}{\exp{(\hbar\omega'/k_B T')}-1}&=&\gamma^2(1+\beta\cos\theta)^2 \frac{{\omega}^2}{4\pi^3c^3}\frac{1}{\exp{(\hbar{\omega}/k_BT)}-1}\\
\nonumber
\frac{1}{\exp{(\hbar\omega'/k_BT')}-1}&=& \frac{1}{\exp{(\hbar{\omega}/k_BT)}-1}\\
T'&=&\frac{T}{\gamma(1+\beta\cos\theta)}\approx\frac{T}{(1+\beta\cos\theta)},
\end{eqnarray}
where $T$ is the temperature of the stationary blackbody source. The temperature of a stationary black body source as seen by a moving object appears as a different temperature to the stationary source. 
\section{Radiation damping and thermalisation of a perfectly reflecting mirror outside a 3D cavity}
\label{supA}
Consider a small perfectly reflecting disk of area $A$ located outside a blackbody source. Photons emanating from a hole in the wall of a 3-D blackbody cavity illuminate the disk. The disk is in motion with a velocity $v_z$ along the $z~$axis. It's surface is normal to the $z$ axis. We now work in the moving frame and drop the dashed nomenclature for this frame such that photons now travel at an angle $\theta$ in this frame delivering a momentum $p=2\hbar k\cos\theta=2\hbar \omega\cos\theta/c$ to the disk. The photons from the BB source trace out volume $c A\cos\theta$ per unit time. The total number of photons per unit time and per solid angle per angular frequency from eq A2 by $dN/d\omega d\Omega dt=\rho_n  A c |\cos \theta|$. 
Dropping the ' such that $\rho_n' \rightarrow \rho_n=\frac{\omega^2}{4\pi^3 c^3}\frac{1}{\exp{[\hbar\omega(1+\beta \cos \theta)/k_B T]}-1}$, $T$ is the temperature in the stationary frame.
The total force on the disk is therefore given by the product of the momentum change per photon, $dN /d\omega d\Omega dt$ integrated over the solid angle $d\Omega$ and the angular frequency $d\omega$. To calculate the maximum damping force we integrate over the half sphere ($0\le\theta\le \pi/2$) and ($0\le\phi\le 2\pi$)
\begin{eqnarray}
\nonumber
F_z&=&\int_0^{\pi/2}\int_0^{2\pi}\int_{0}^{\infty}{\rho_n cA \cos \theta ~2\hbar \frac{\omega}{c} \cos{\theta}~d\Omega d\omega}\\
\nonumber
&=&\int_0^{\pi/2}\int_0^{2\pi}\int_{0}^{\infty}\frac{\hbar \omega^3 }{4\pi^3c^3} \frac{A\cos^2\theta}{\exp{[\hbar\omega (1 + \beta \cos \theta)/k_BT]}-1} \sin \theta d\theta d\phi d\omega\\
\nonumber
&\approx&\int_0^{\pi/2}\int_0^{2\pi}\Biggl[\frac{\pi Ak_B^4T^4\cos^2\theta}{30c^3\hbar^3}-\frac{2\pi Ak_B^4T^4\cos^3\theta}{15c^3\hbar^3}\beta\Biggr]\sin\theta d\theta d\phi\\
\nonumber
&=&\frac{A\pi^2k_B^4T^4}{45c^3\hbar^3}-\frac{A\pi^2k_B^4T^4}{15Mc^4\hbar^3}~M v_z\\
&=&\frac{A\pi^2k_B^4T^4}{45c^3\hbar^3}-\Gamma_z~Mv_z,
\label{aeqn1}
\end{eqnarray}
 where $\Gamma_z=\frac{A\pi^2 k_B^4T^4}{15Mc^4\hbar^3}$ and $M$ is the mass of the perfectly reflecting disk/mirror. The first term in equation (\ref{aeqn1}) is the usual radiation pressure force while the second term is the radiation damping. As the light has thermal fluctuations, they lead to thermalisation of the disk to the light \cite{Einstein1909,Mansuripur2017}. The fluctuation in the photon number per unit volume per unit solid angle per unit angular frequency from the blackbody is given by the usual value (per volume per angular frequency) divided by $4 \pi$ to be $\Delta N^2=\frac{\omega^2}{4\pi^3c^3}~\frac{\exp{(\hbar\omega/k_BT)}}{[\exp{(\hbar\omega/k_BT)}-1]^2}$. On reflection, the energy delivered by each photon to the disk is $p^2/2M$. The gain in energy per unit time is then given by the product of variance per unit volume per unit solid angle per angular frequency multiplied by the volume travelled by photons at angle $\theta$ per second as $A c \cos{\theta}$. This is finally multiplied by the square of the momentum change ($p=2\hbar k\cos\theta=2\hbar \omega\cos\theta/c$ ) per photon on reflection. We integrate this expression over the solid angle ($2\pi$) received by the reflective disk and is given by
\begin{eqnarray}
\nonumber
\Delta \dot{E}&=&\frac{1}{2M}\int_0^{\pi/2}\int_0^{2\pi}\int_{0}^{\infty}{cA\cos\theta~ \Delta N^2~\frac{4\hbar^2\omega^2\cos^2\theta}{c^2}~\sin\theta d\theta d\phi d\omega}\\
\nonumber
&=&\frac{A}{2M}\int_0^{\pi/2}\int_0^{2\pi}~\int_{0}^{\infty}{~\frac{\omega^2}{4\pi^3 c^2}~\frac{\exp{[\hbar\omega/k_BT]}}{[\exp{\hbar\omega/k_BT}-1]^2}~\frac{4\hbar^2\omega^2\cos^3\theta}{c^2}~\sin\theta d\theta d\phi d\omega}\\
&=&\frac{A\pi^2k_B^5T^5}{15M\hbar^3c^4}
\label{aeqn2}
\end{eqnarray}
\\
At equilibrium, via equipartition, we have $v_z^2=k_BT_{cm}/M$. In addition, the energy loss rate or power loss, is given by the product of the damping force and the velocity. In equilibrium this must equal the power increase $\Delta \dot{E}$ such that
\begin{eqnarray}
\nonumber
(\Gamma M v_z). v_z&=&\Delta \dot{E}\\
\nonumber
\frac{A\pi^3k_B^4T^4}{15Mc^4\hbar^3}~k_B T_{cm}&=&\frac{A\pi^3k_B^5T^5}{15M\hbar^3c^4}\\
T_{cm}&=&T.
\label{aeqn30}
\end{eqnarray}
This important result is that at equilibrium the centre-of-mass frequency is equal to the black body temperature due to radiation damping from thermal photon gas. 

\section{ Radiation damping of a perfectly reflective disk using thermal light from a 2D micro-cavity} 
\label{supD}
\textbf{The energy density of a 2D cavity:} Thermal light with a well defined chemical potential has been realised in both 2D and 1D. For the 2D case, light is thermalised in the transverse modes of an optical cavity consisting of single longitudinal mode $q$\cite{KlaersNatPhys2010,Sobyanin2013}. The energy of the photons in this type of source is given by $\hbar\omega_c+(n_x+n_y+1)\hbar \Omega$, where $n_x$ and $n_y$ are the transverse mode numbers along the $x-$axis and $y-$axis. The cut-off frequency is given by $\omega_c=q c\pi/D_0$ and the transverse mode separation is $\Omega=2 \pi c/n\sqrt{D_0R/2}$, where, $D_0$ is the separation between the cavity mirrors and $R$ is the radius of curvature of the cavity  mirrors. The degeneracy of transverse modes is $2(n_x+n_y+1)$, where the factor $2$ accounts for the two polarization states. The energy density per mode inside such a cavity can be expressed as \cite{MullerEPRA2019}
\begin{eqnarray}
\bar{u}&=&\frac{1}{V_R}\sum_{n_x=0}^{\infty}\sum_{n_y=0}^{\infty}2(n_x+n_y+1)~\frac{\hbar(\omega_c+(n_x+n_y+1)\Omega) }{\exp{[(\hbar(\omega_c+(n_x+n_y+1)\Omega)-\mu_c) /k_BT]}-1},
\label{eqnD1}
\end{eqnarray}
where $V_R$ is the volume of the cavity. In the continuum limit, which has been shown to be well approximated by this cavity, $\Omega \to 0$. In this limit Eq. (\ref{eqnD1}) is the energy density per unit angular frequency given by \cite{MullerEPRA2019}
\begin{eqnarray}
\bar{u}(\omega)&=& \frac{A_R}{4\pi V_R}~\frac{\omega }{c^2} \frac{\hbar(\omega_c+\omega) }{\exp{[(\hbar(\omega_c+\omega)-\mu_c) /k_BT]}-1},
\label{eqnD2}
\end{eqnarray}
where $A_R$ is the surface area of the cavity. Assuming that the cavity mirror is a spherical cap, one can express $V_R=\pi D_0^2(3R-D_0/2)/6\approx \pi D_0^2R/2$ and $A_R=2\pi R D_0$, where we have used the fact $R\gg D_0$  \cite{MullerEPRA2019}. This means that we have $A_R/4\pi V_R=1/\pi D_0=\omega_c/q  c \pi^2$.  With these substitutions, Eq. (\ref{eqnD2}) is expressed as
\begin{eqnarray}
\bar{u}(\omega)&=&\frac{1}{q}\frac{\omega_c \omega }{\pi^2 c^3} \frac{\hbar(\omega_c+\omega) }{\exp{[(\hbar(\omega_c+\omega)-\mu_c) /k_BT]}-1}.
\label{eqnD4}
\end{eqnarray}
Except for the prefactor ($1/q$), equation (\ref{eqnD4}) has a similar form to the 3D blackbody radiation density \cite{Mansuripur2017} e.g. $\bar{u}(\omega)=\frac{\omega^2}{\pi^2c^3}\frac{\hbar\omega}{\exp{(\hbar\omega/k_BT)}-1}$ and all the Lorentz transformations presented in Section \ref{supA0} can be applied.

\textbf{Photon statistics:} For a particular angular frequency, the average number of photons in the cavity can be found by multiplying the photon number density $\bar{u}/\hbar (\omega_c+\omega)$ by the volume of the cavity $V_R$. The rate at which photons escape from the 2D cavity is determined by the mean photon lifetime ($\tau=\frac{2 n D_0}{c T_r} $) multiplied by the number of photons inside the cavity, where $T_r$ is the cavity transmission. Finally, assuming emission into $2 \pi$ solid angle, the average rate of escape of photons per unit solid angle and per angular frequency is,    
\begin{eqnarray}
\nonumber
\dot{\bar{N}}&=&\frac{1}{2\pi}\frac{cV_RT_r}{2 n D_0}  \frac{\bar{u}}{\hbar(\omega_c+\omega)}\\
\nonumber
&=&\frac{cV_RT_r}{n q D_0}   \frac{\omega_c \omega }{4\pi^3 c^3} \frac{1 }{\exp{[(\hbar(\omega_c+\omega)-\mu_c) /k_BT]}-1}\\
&=&\frac{V_RT_r\exp{[\mu_c/k_BT]}}{q n D_0}  \frac{\omega_c \omega }{4\pi^3 c^2} \frac{1 }{\exp{[\hbar(\omega_c+\omega) /k_BT]}},
\end{eqnarray}
where in the last equation we have used the fact that for $\hbar(\omega_c+\omega)> \mu_c) \gg k_BT$, $\exp{[(\hbar(\omega_c+\omega)-\mu_c) /k_BT]}\gg 1$. Furthermore, we have separated the chemical potential term which acts as an amplification factor \cite{Wurfel_1982}. The associated variance of the photon number per unit volume per solid angle per unit angular frequency inside the cavity is 
\begin{eqnarray}
\Delta N^2&=&\frac{V_RT_r}{n q D_0}\frac{\omega_c \omega }{4 \pi^3 c^2} \frac{1}{\exp{[(\hbar(\omega_c+\omega)-\mu_c) /k_BT]}}
\end{eqnarray}

\textbf{The output power of a 2D cavity:} The output power of such a cavity can be found as
\begin{eqnarray}
\nonumber
P&=& \int_0^\infty T_r\frac{c}{2D_0} \bar{u}V_R d\omega\\
\nonumber
&=&\int_0^\infty \frac{cV_RT_r}{2qnD_0}  \frac{\omega_c \omega }{\pi^2 c^3} \frac{\hbar(\omega_c+\omega) ~d\omega }{\exp{[(\hbar(\omega_c+\omega)-\mu_c) /k_BT]}-1}\\
&=&\exp{\big[ (\mu_c-\hbar\omega_c)/k_BT\big]}~\frac{V_rT_r\omega_ck_B^2T^2(2k_BT+\hbar\omega_c)}{2qnD_0\pi^2\hbar^2c^2}
\end{eqnarray}
For the experimentally realized $2D$ photon gas and the associated cavity parameters \cite{KlaersNatPhys2010} e.g. $R=1~$m, $D_0=1.46~\mu m$, $q=7$, $T_r=1.5\times 10^{-5}$, we get $P\approx 22~$nW. The experimentally measured power was $\approx 50~$nW. This discrepancy can arise from a number of factors including the cavity transmission co-efficient, the cavity length and the cavity mirror diameters.

\textbf{Radiation damping:} In calculating the radiation damping that a perfectly  reflecting disk encounters we assume that all the light coming out from the cavity is captured by a lens (see Fig. \ref{fig0} in the main text) and amplified by by gain $G$. The linear momentum of a photon is $p=\hbar(\omega_c+\omega)/c$. On reflection from a stationary object each photon delivers a momentum of $2p\cos\theta$ in the z-direction. This is the usual radiation pressure force and is given by the product of the number of photons emitted per unit time per unit solid angle per unit angular frequency $\dot{\bar{N}}$ multiplied by $2p\cos\theta$ and gain $G$ integrated over the solid angle (2 $\pi$) and the angular frequency. As before due to the motion of the mirror a velocity dependent force exists and the mirror encounters a drag force. The mirror is allowed to move along the $z-$axis with a velocity $v_z$ and using the appropriate transformations of supplementary Section \ref{supA0}, and assuming that the chemical potential is the same in each frame, the force is given by 
\begin{eqnarray}
\nonumber
F_z&=& \frac{G V_RT_r \exp{[\mu_c/k_BT]}}{qnD_0}\int_0^{\pi/2}\int_0^{2\pi} \int_0^{\infty}{ \frac{\omega_c \omega }{4\pi^3 c^3} \frac{2\hbar(\omega_c+\omega)\cos^2\theta \sin\theta d\theta d\phi d\omega }{\exp{[\hbar(\omega_c+\omega)(1+\beta\cos \theta) /k_BT]}}}\\
\nonumber
&\approx&\frac{G V_RT_r \exp{[(\mu_c-\hbar\omega_c)/k_BT]}}{3qnD_0}  \frac{\omega_ck_B^2T^2 (2k_BT+\hbar\omega_c)}{\pi^2 \hbar^2c^3}\\
\nonumber
&&-\frac{G V_RT_r \exp{[(\mu_c-\hbar\omega_c)/k_BT]}}{4qnD_0}  \frac{\omega_ck_BT (6k_B^2T^2+4\hbar\omega_ck_BT-\hbar^2\omega_c^2)}{\pi^2 \hbar^2c^3}\beta_z\\
&\approx &\frac{G V_RT_r }{3qnD_0}  \frac{\exp{[(\mu_c-\hbar\omega_c)/k_BT]}\hbar^2\omega_c^2k_B^2T^2}{\pi^2 \hbar^3 c^3}-\frac{G V_RT_r }{4qnD_0}  \frac{\exp{[(\mu_c-\hbar\omega_c)/k_BT]}\hbar^3\omega_c^3k_BT }{\pi^2 \hbar^3c^3}\beta_z
\label{beqn8}
\end{eqnarray}
where in the second equation, we have used a Taylor series in $\beta_z$ and kept only first order. Furthermore, in deriving the final result (Eq. (\ref{beqn8})), we have used the fact that $\hbar\omega_c \gg k_BT$. The difference between the exact (the 2nd equation) and the approximate (the last equation of (\ref{beqn8}) is less than $10\%$. Note that the first term in Eq. (\ref{beqn8}) represents the usual radiation pressure force while the $2^{nd}$ term is the drag force. Finally, the damping rate is 
 \begin{eqnarray}
 \Gamma_z&=&\frac{G V_RT_r }{4qnD_0M}  \frac{\exp{[(\mu_c-\hbar\omega_c)/k_BT]}\hbar^3\omega_c^3k_BT }{\pi^2 \hbar^3c^4},
 \label{beqn2}
 \end{eqnarray}
 where we have used $\beta=v/c$ and $M$ is the mass of the disc. This is the result we have shown in the main article Eq. (\ref{eqn4}).
 
 The corresponding rate of gain in energy due to the fluctuating momentum associated with the photon number fluctuation is
 \begin{eqnarray}
 \nonumber
 \Delta \dot{E}&=&\frac{1}{2M}\int_0^{\pi/2}\int_0^{2\pi}\int_0^{\infty}{G\Delta N^2 \frac{4\hbar^2(\omega_c+\omega)^2}{c^2}}\cos^2 \theta\cos\theta\sin\theta d\theta d\phi d\omega\\
 \nonumber
 &=& \frac{G V_RT_r\exp{[\mu_c/k_BT]}}{2q nD_0M} \int_0^{\pi/2}\int_0^{2\pi}\int_0^{\infty}\frac{\omega_c \omega }{4\pi^3 c^4} \frac{4\hbar^2(\omega_c+\omega)^2 \cos^2 \theta\cos\theta\sin\theta d\theta d\phi  d\omega}{\exp{[(\hbar(\omega_c+\omega)) /k_BT]}}\\
 \nonumber
 &=&\frac{G V_RT_r \exp{[(\mu_c-\hbar\omega_c)/k_BT]}}{qnD_0M}  \frac{\omega_ck_B^2T^2 (6k_B^2T^2+4\hbar\omega_ck_BT+\hbar^2\omega_c^2)}{4\pi^2 \hbar^2c^4}\\
 &\approx&\frac{G V_RT_r \exp{[(\mu_c-\hbar\omega_c)/k_BT]}}{4qnD_0M}  \frac{k_B^2T^2 \hbar^3\omega_c^3}{\pi^2 \hbar^3c^4}
 \end{eqnarray}
In deriving the last expression we assume that $\hbar\omega_c >\mu_c\gg k_BT \gg \hbar\Omega$. Now for the centre of mass temperature we have $(M\Gamma_z v_z^2=\Delta \dot{E}$ and therefore $T_{cm}=T$, where we have used $v_z^2=k_BT_{cm}/M$.
\section{A perfectly reflecting mirror and a laser beam}
\label{supE}
Let us consider a laser beam  consists of $N$ photons and of frequency $\omega_c$ is incident on the perfectly reflecting mirror considered above. The momentum $p$ of each photon is $\hbar \omega_c/c$. In this case the force that the incoming photons exert on the mirror including the Doppler effect is 
\begin{eqnarray}
\nonumber
F_z&=& N~2\hbar \omega_c(1+\beta_z)/c\\
\nonumber
&=&\frac{2N\hbar\omega_c}{c}+\frac{2N\hbar\omega_c v_z}{c^2}.
\end{eqnarray}
The damping rate is $\Gamma_z=\frac{2N \hbar\omega_c}{Mc^2}$, where $M$ is the mass of the mirror.
The variance of the photon number of a Poisson distributed laser beam is the same as the average photon number $N$ in the beam. Now the gain in energy due to the fluctuation in the photon number is 
\begin{eqnarray}
\nonumber
\Delta \dot{E}&=&\frac{1}{2M} N~(2p)^2\\
&=& \frac{2N \hbar^2\omega_c^2}{Mc^2}
\end{eqnarray}
Now the centre-of-mass temperature of the mirror is 
\begin{eqnarray}
\nonumber
\Gamma M v_z\hat{z}.v_z\hat{z}&=&\Delta E\\
T_{cm}&=&\frac{\hbar\omega_l}{k_B}
\end{eqnarray}
\section{Thermalized photon gas and a levitated dielectric sphere}
\label{supC}
Here, we consider a levitated dielectric sphere whose radius is much smaller than the wavelength of light such that $r\ll 2\pi c/\omega_c$. The sphere is illuminated by thermal light from a 2D cavity and has a scattering cross section $\sigma_s=\frac{\alpha^2\omega_i^4}{6\pi\epsilon_0^2c^4}$ \cite{BohrenHuffman2007}, where $\alpha$ is its polarizability and $\omega_i=\omega_c+\omega$ is the angular frequency of the incident light. We approximate that polarizability is approximately constant over the spectral range of the incident photons and  $\sigma_s$ is independent of velocity. We assume that the incoming light covering a solid angle of $2\pi$ steradians ($\theta_i^{max}=\pi/2$ and $\phi_i^{max}=2\pi$) is focused to a spot of area $A_w$ using a lens. Furthermore, we assume that the particle is at the focus of the thermal light  and that the average number of photons per unit time that interact with the particle of cross-section $\sigma_s$ is


\begin{eqnarray}
\frac{dN}{dt}=&  \frac{\sigma_s}{A_w}\frac{G V_RT_r\exp{[\mu_c/k_BT]}}{q n D_0} \frac{\omega_c \omega }{4\pi^3 c^2} \frac{1 }{\exp{[\hbar(\omega_c+\omega) /k_BT]}}.
\end{eqnarray}
The wavevector of the incident light is $\mathbf{k_}_i=[k_x~k_y~k_z]$ with $k_i=\omega_i/c$. The particle is free to move along all three axes with a velocity $\mathbf{v}=[v_x~v_y~v_z]$. Owing to the unpolarized nature of the thermal incident photons, scattered photons are isotropically distributed in all $4 \pi$ steradian \cite{ItanoWinelandPRA82}. The scattered photon thermal distribution remains the same. In the reference frame of the particle, the frequency of an incident and a scattered photon is $\omega_i(1+\beta)$, where $\beta=\frac{\mathbf{v}\cdot\mathbf{k}_i}{ck_i}$. After a scattering event, each photon delivers a momentum equivalent to $\mathbf{p}=\hbar k_i(1+\beta)(\mathbf{\Theta}_i-\mathbf{\Theta}_s)$, where $\mathbf{\Theta}_i=[\sin\theta_i\cos\phi_i~\sin\theta_i\sin\phi_i~\cos\theta_i]$ and $\mathbf{\Theta}_s=[\sin\theta_s\cos\phi_s~\sin\theta_s\sin\phi_s~\cos\theta_s]$. Here, $\theta_i$ and $\phi_i$ represent the polar and the azimuthal angles that the wavevector of an incoming photon makes with the $-z~$axis and the $+x~$axis, respectively. $\theta_s$ and $\phi_s$ are the corresponding angles that the wavevector of a scattered photons makes with the same axes. With the appropriate Lorentz transformation shown above, forces along the three axes that all photons exert are


\begin{eqnarray}
\nonumber
\mathbf{F}&=&\frac{ GV_RT_r}{q nD_0}\frac{1}{4\pi}\int_0^{\theta_s^{max}}\int_0^{\phi_s^{max}}\int_0^{\theta_i^{max}}\int_0^{\phi_i^{max}}\int_{0}^{\infty}\Bigg[\frac{\sigma_{s}}{A_w}\frac{\omega_c\omega}{4\pi^3c^2}~ \frac{\exp{(\mu_c/k_BT)}}{\exp{\Bigl[\hbar\omega_i(1+\beta)/k_BT\Bigr]}} \hbar k_i\Bigl(\mathbf{\Theta}_i -\mathbf{\Theta}_s\Bigr) \Bigg]d\omega d\Omega_i~d\Omega_s \\
&\approx&\frac{GV_RT_r}{qnA_wD_0}~\frac{\exp{[(\mu_c-\hbar\omega_c)/k_BT]}\alpha^2k_B^2T^2\omega_c^6}{24\pi^3\hbar \epsilon_0^2c^7}~\mathbf{\Upsilon}- \frac{GV_RT_r}{qnA_wD_0}~\frac{\exp{[(\mu_c-\hbar\omega_c)/k_BT]}\alpha^2k_BT\omega_c^7}{36\pi^3 \epsilon_0^2c^8}~\mathbf{v},
\end{eqnarray}
where $\mathbf{\Upsilon}=[0~0~1]$, $d\Omega_i=\sin\theta_i  d\phi_i d\theta_i$ with $\theta_i^{max}=\pi/2$, $\phi_i^{max}=2\pi$, and $d\Omega_s=\sin\theta_s  d\phi_s d\theta_s$ with $\theta_s^{max}=\pi$ and $\phi_s^{max}=2\pi$. As before we have Taylor series expanded the velocity dependent terms around $\beta_{x,y,z}\approx 0$ and kept only first order terms. We have also used the fact that $\hbar\omega_c\gg k_BT$. The first term in the equation above represents the radiation pressure while the second term is the drag force. The damping rate for the all three axes can be represented as 
\begin{eqnarray}
\Gamma_{x,y,z}=\frac{GV_RT_r}{qnA_wD_0M}~\frac{\exp{[(\mu_c-\hbar\omega_c)/k_BT]}\alpha^2k_BT\omega_c^7}{36\pi^3 \epsilon_0^2c^8}.
\end{eqnarray}

The gain in energy due to a single scattering event, ignoring the Doppler effect, is $\frac{p^2}{2M}$, where $M$ is the mass of the particle as before. Now the rate in gain in energy for the thermal light source is

\begin{eqnarray}
\nonumber
\Delta \dot E&=&\frac{1}{2M}\frac{ GV_RT_r}{q nD_0}\frac{1}{4\pi}\int_0^{\theta_s^{max}}\int_0^{\phi_s^{max}}\int_0^{\theta_i^{max}}\int_0^{\phi_i^{max}}\int_{0}^{\infty}\Bigg[ \frac{\sigma_s}{A_w}\frac{\omega_c\omega}{4\pi^3c^2}~ \frac{\exp{(\mu_c/k_BT)}}{\exp{\Bigl[\hbar\omega_i/k_BT\Bigr]}} \hbar^2 k_i^2\Bigl(\mathbf{\Theta}_i -\mathbf{\Theta}_s\Bigr)^2 \Bigg]d\omega d\Omega_i~d\Omega_s \\
&\approx& \mathbf{\Lambda} \frac{GV_RT_r}{qnA_wD_0M}~\frac{\exp{[(\mu_c-\hbar\omega_c)/k_BT]}\alpha^2k_B^2T^2\omega_c^7}{36\pi^3 \epsilon_0^2c^8},
\end{eqnarray}
\normalsize
where $\mathbf{\Lambda}=[1~1~1]$. As before in deriving the final result we have used the fact that $\hbar\omega_c\gg k_BT$.

\textbf{Equilibrium centre-of-mass temperature:} At equilibrium we have $v_{x,y,z}^2=k_BT_{cm}/M$ and the loss and the gain in energy must be balanced
\begin{eqnarray}
\nonumber
M \Gamma \mathbf{v}\cdot \mathbf{v}&=& \Delta \dot E\\
T_{cm}&=& [1~1~1] T
\label{aeqn3}
\end{eqnarray}

\end{document}